\documentclass[twocolumn,superscriptaddress]{revtex4-2}

\usepackage{graphicx}
\usepackage{bm}
\usepackage{xcolor}
\usepackage{amsmath}

\usepackage{hyperref}
\hypersetup{
    colorlinks=true,
    linkcolor=blue,
    filecolor=blue,      
    urlcolor=blue,
    pdftitle={main},
    pdfpagemode=FullScreen,
    }
\begin{document}


\title{Finite-Element Simulations of Rotating Neutron Stars with Anisotropic Crusts and Continuous Gravitational Waves}

\author{J.A. Morales}
\email{jormoral@iu.edu}
\affiliation{ Center for the Exploration of Energy and Matter and Department of Physics, Indiana University, Bloomington, IN 47405, USA }

\author{C.J. Horowitz}
 \email{horowit@iu.edu}
\affiliation{ Center for the Exploration of Energy and Matter and Department of Physics, Indiana University, Bloomington, IN 47405, USA }

\date{\today}

\begin{abstract}
``Mountains'', or non-axisymmetrical deformations in the elastic crust of rotating neutron stars are efficient radiators of continuous gravitational waves. Recently, small anisotropies were observed in the solid innermost inner core of the Earth \cite{Phdam2023}. We implement three-dimensional finite-element simulations to study mountains sourced by modest anisotropies in the solid crust of rotating neutron stars. We find that anisotropic mountains may be detectable by current ground-based gravitational-wave detectors and might explain several observed phenomena, confirming the results of a previous work on less realistic neutron star models \cite{Morales2023}. In particular, we find that a slightly anisotropic neutron star crust that changes its rotation rate modestly can support an ellipticity of a few $\times$ 10$^{-9}$, which is equivalent to the upper bounds on the ellipticity of some nearby and rapidlly spinning pulsars \cite{Abbott2022_2}. 
\end{abstract}


\maketitle



\section{Introduction \label{sec:Intro}}

In recent years, the scientific community has been profoundly impacted by groundbreaking discoveries regarding gravitational waves (GWs). The first detection of GWs occurred on September 14, 2015, when the LIGO and VIRGO collaborations identified bursts of GWs resulting from the inspiral and merger of two black holes \cite{Abbott2016}. Approximately two years later, on August 17, 2017, these collaborations made another historic observation: they detected GWs from the inspiral and merger of two neutron stars (NSs) \cite{Abbott2017}. Additionally, several pulsar-timing-array (PTA) collaborations have announced varying levels of confidence with respect to evidence for the existence of nano-Hertz stochastic GWs \cite{Agazie2023,Antoniadis2023,Xu2023,Reardon2023},.

However, there is another type of GWs that has continually eluded detection: continuous GWs (CGWs). CGWs are continuous trains of GWs with a quasi-monochromatic frequency that evolve slowly over time. The most promising sources of CGWs are individual spinning NSs with time-varying mass quadrupoles. The two primary time-varying mass quadrupoles that can generate CGWs in rotating NSs are r-mode instabilities \cite{Andersson2023,Gittin2023} and non-axisymmetric deformations sustained by magnetic and/or elastic forces \cite{Gittins2024,Hutchins2023,Rossetto2023}. r-mode instabilities can be suppressed by various mechanisms. For instance, at low temperatures, shear viscosity can suppress the r-mode instability before it grows to a detectable level, while at high temperatures, bulk viscosity can have the same effect \cite{Shirke2023,Haskell2015,Alford2012}. In addition, mode coupling can also suppress the r-mode instability \cite{Bondarescu2009}. Sustained non-axisymmetric deformations in rotating NSs are commonly referred to as ``mountains''. Mountains sustained by magnetic forces are likely to emit CGWs that fall below the frequency range where ground-based detectors are most sensitive \cite{Pagliaro2023}. This is because the strong magnetic fields that create these mountains may also exert a magnetic torque that rapidly spins down the NS. Conversely, mountains can form in the crust (the solid outermost kilometer of a typical NS) and be sustained mainly by elastic forces if the magnetic fields are weak. These mountains are typically known as ``elastic mountains'' and are the most promising sources of CGWs because there is no significant magnetic torque to rapidly spin down the star and shift the frequency of the emitted CGWs to less sensitive bands of ground-based GW detectors. This work is focused on elastic mountains.

The instrumentation of ground-based GW detectors and the data analysis techniques used in CGW observational campaigns are continuously improving \cite{Covas2024,Ashok2024,Jaume2024,Sauter2024,Miller2024}. Remarkably, researchers started to develop neural network architectures to perform observational campaigns of CGWs \cite{Joshi2023}. Additionally, more galactic pulsars are being discovered, and more electromagnetic information is being collected from potential CGW sources \cite{Padmanabh2024,Turner2024}. These advances enhance our GW constraints on known CGW sources and increase the likelihood of the first CGW detection in the coming years. However, there has not yet been a CGW detection, although there has been significant progress in the upper limits on the strengths of CGW signals. \cite{Owen2024,Steltner2023,Ming2022,Abbott2022,Abbott2022_2}.

Given the lack of CGW detections, elastic mountains remain poorly understood. Nonetheless, there has been significant progress in the theoretical understanding of elastic mountains over the years. The calculation of the breaking strain of the NS crust has enabled the determination of the maximum mountain that the crust can support \cite{Caplan2018,Horowitz2009,Gittins2023_2,Morales2022}. Additionally, research studies have investigated elastic mountains arising from non-axisymmetric formation channels, including electron capture reactions \cite{Ushomirsky2000} and temperature variations \cite{Hutchins2023}. 

Crystals are anisotropic and tend to be oriented by sedimentation, freezing, recrystallization, deformation, and flow. The NS crust may be anisotropic as accretion moves material to greater densities where it may melt or freeze. Furthermore, recent observations of seismic waves reverberating through the Earth's center  showed that the solid innermost inner core of the Earth is anisotropic \cite{Phdam2023}, with material properties that depend on direction. Motivated by these observations, in an earlier publication \cite{Morales2023} we considered a solid NS crust with anisotropic stresses. Our study used a simplified two-dimensional geometry and a constant mass density crust model.  It suggested that the centrifugal stress can form detectable mountains when coupled with the anisotropic shear modulus of the crust. We found that the size of these mountains evolves with the difference between the square of the final and initial rotational frequencies and scales linearly with the degree of anisotropy within the crustal material. 

In this research work, we use three-dimensional finite-element simulations of a more realistic NS to enhance our model mountains that grow as the star with an anisotropic crust spins up or down. We simulate how the anisotropic shear modulus deforms the star as the crust is centrifugally stressed by a change in rotation rate. We also discuss the gravitational radiation emitted by this kind of deformations.

 We emphasize that the basic assumption in this project is that the elastic stress of the crust is very slightly anisotropic, even when averaged over the whole crust. Specifically, we assume that the elastic stress tensor is given by, 
\begin{equation}
    t_{ij} = 2 \mu \bar{\varepsilon}_{ij} + 2 \mu \langle \psi \rangle \bar{\varepsilon}_{xx} \delta_{ix} \delta_{jx} \ ,
    \label{eqn:elastic_stress_0}
\end{equation}
where the first term is the contribution from the isotropic angle-averaged shear modulus, while the second term is the contribution from a small anisotropic perturbation to the isotropic angle-averaged shear modulus. The elastic stress tensor defined in equation (\ref{eqn:elastic_stress_0}) is explained with detail in section \ref{subsec:crust}.

This paper is organized as follows. In section \ref{sec:ns_mountains}, we introduce the concept of NS mountains. In section \ref{sec:background}, we present the background NS against which we perform a perturbative analysis to study the anisotropic NS crust. In section \ref{sec:perturbations}, we describe the perturbations in the fluid core and elastic crust of a NS. In section \ref{sec:methods}, we describe the numerical methods that we use to perform our analysis, including the finite-element method. In section \ref{sec:results}, we talk about the results and discuss some of their implications, including the ellipticity that a spun-up (-down) NS with an anisotropic crust can support. Finally, in section \ref{sec:conclusions}, we present some concluding remarks.  


\section{Neutron Star Mountains \label{sec:ns_mountains}}

The size of NS mountains is quantified by the ellipticity. Qualitatively, the ellipticity measures the degree of non-axisymmetrical deviation from perfect sphericity. The equatorial ellipticity is defined as
\begin{equation}
    e\equiv \frac{I_{xx} - I_{yy}}{I_{zz}} \ ,
    \label{eqn:equatorial_ellipticity}
\end{equation}
where $I_{zz}$ is the moment of inertia about the rotation axis, and $I_{xx}$ and $I_{yy}$ are moments of inertia about the x and y axes, respectively.

The CGWs emitted by a rotating deformed NS have a characteristic strain amplitude
\begin{equation}
    h_0 = \frac{16 \pi^2 G}{c^4} \frac{I_{zz}}{d} \Omega^2 e\ ,
    \label{eqn:intrinsic_strain}
\end{equation}
where $d$ and $\Omega$ are the distance and the rotational frequency of the deformed and spinning NS. An axisymmetric NS does not emit CGWs because $I_{xx}=I_{yy}$, which implies a vanishing ellipticity.

Horowitz and Kadau \cite{Horowitz2009} performed molecular dynamics simulations and found that the breaking strain of the NS star crust is as large as 0.1 because the crust is under great pressure, which prevents the formation of voids and fractures. As a consequence, the maximum ellipticity that the crust of a traditional NS can support is $e_{max} \sim 10^{-6}$ \cite{Gittins2024,Morales2022,Ushomirsky2000}. Numerous CGW searches for known or suspected NSs have found upper limits on the ellipticity that are smaller than $e_{max}$. Examples include references \cite{Ming2022} and \cite{Abbott2022}. On the other hand, the most constraining upper limits on the ellipticity are $e_{min} \sim 10^{-9}$ for some close-by and rapidly-spinning pulsars \cite{Abbott2022}. Therefore, ellipticities that lie between $e_{max}$ and $e_{min}$ are very interesting from an observational standpoint. In the following sections, we confirm what we found in our previous study: small anisotropies can lead to the formation and evolution of interesting ellipticities that have values between $e_{max}$ and $e_{min}$ and might be detectable by current ground-based GW detectors \cite{Morales2023}.


\section{The Background Star \label{sec:background}}

We consider a non-rotating, canonical NS with a mass of 1.4 M$_{\odot}$ and a radius of 10 km as our background star. The matter within this NS is a cold perfect fluid characterized by mass density $\rho$ and barotropic pressure $p$. We use a polytropic equation of state (EOS)
\begin{equation}
p(\rho) = K_\text{p} \rho^{1 + 1/n}
\label{eqn:EOS}
\end{equation}
to describe the background NS. This EOS effectively models massive NSs. Here, $n = 1$ is the polytropic index and $K_\text{p}$ is the proportionality constant, which is adjusted to ensure that the NS has a radius of 10 km when its mass is 1.4 $M_{\odot}$. In the slow-rotation approximation, the effects of the centrifugal force are much smaller than the effects of pressure and gravity. Therefore, it is sensible to obtain the Newtonian structure of the NS neglecting rotation (see section \ref{sec:perturbations}). The Newtonian structure equations of mechanical equilibrium in the fluid background star are the following:
\begin{equation}
m^{\prime} = 4 \pi r^2 \rho \label{eqn:SEa}
\end{equation}
\begin{equation}
p^{\prime} = -\rho \Phi^{\prime} \label{eqn:SEb}
\end{equation}
\begin{equation}
\Phi^{\prime} = \frac{Gm}{r^2} \label{eqn:SEc} \ .
\end{equation}
In these equations, $m(r)$ is the enclosed mass and $\Phi(r)$ is the gravitational potential. The prime represents radial derivatives. Essentially, the pressure from the electrons and the nuclei balance gravity. In Newtonian mechanics, these equations give the exact background structure of a NS that is spinning up from rest. On the other hand, if the NS is initially spinning with an angular frequency of $\omega_0$, equations (\ref{eqn:SEa}-\ref{eqn:SEc}) give a reasonable estimate of the real structure of the NS as long as $\omega_0^2 \ll \omega_k^2$, where $\omega_k$ is the Kepler frequency. For the present investigation, equations (\ref{eqn:SEa}-\ref{eqn:SEc}) suffice. The most general form of equations (\ref{eqn:SEa}-\ref{eqn:SEc}) for a Newtonian star under rotation is given by the equations
\begin{equation}
    \vec{\nabla} p + \rho \vec{\nabla} \Phi = \vec{f_0} \ ,
    \label{eqn:SE_with_rotation_a}
\end{equation}
\begin{equation}
    \nabla^2 \Phi = 4 \pi G \rho \ ,
    \label{eqn:SE_with_rotation_b}
\end{equation}
\begin{equation}
    M = \int_{\Omega_{V}} \rho \ dV \ .
    \label{eqn:SE_with_rotation_c}
\end{equation}
where $\vec{f}_0 = \rho \omega_0^2 s \hat{s}$ is the centrifugal force, $s$ is the distance from the rotation axis, and $\Omega_{V}$ represents the volume of the star. These equations become necessary to obtain the stellar structure for a rapidly rotating NS. Solving equations (\ref{eqn:SE_with_rotation_a}-\ref{eqn:SE_with_rotation_c}) is out of the scope of our investigation. For an explanation on how to solve these equations, see reference \cite{Roxburgh2004}.

In the present study, the NS crust is of paramount importance. We assume that outermost $\sim$ 1 km layer freezes, forming a solid crust in mechanical equilibrium without any elastic stresses. This assumption is explained further in section \ref{sec:perturbations}. Since there are no elastic stresses, the Newtonian structure equations are the same as equations (\ref{eqn:SEa}-\ref{eqn:SEc}). The material parameters of the solid crust are the angle-averaged shear modulus 
\begin{equation} 
\mu(\rho) = \kappa \rho\,
\label{eqn:SM}
\end{equation}
and the bulk modulus
\begin{equation}
K(\rho) = c_s^2 \rho \ .
\label{eqn:BM}
\end{equation}
where $\kappa = 10^{16}$ cm$^2$ s$^{-2}$ \cite{Haskell2015} and $c_s^2 \equiv dp/d \rho$ is the square of the speed of sound. The shear modulus quantifies the resistance of the crustal material to shear, while the bulk modulus quantifies its resistance to changes in volume. We assume the mass densities at the core-crust (bottom) and crust-ocean (top) interfaces are given by $\rho_{bottom} = 2 \times 10^{14}$ g cm$^{-3}$ and $\rho_{top} = 10^{12}$ g cm$^{-3}$. We selected this value for $\rho_{top}$ to avoid numerical difficulties. However, we found that making this value slightly larger or smaller does not affect the results of this work (see Table \ref{tab:simulation_results}).


\section{Equations of Centrifugal Perturbations \label{sec:perturbations}}

As we mentioned in section \ref{sec:background}, we neglect the influence of rotation to obtain the background stellar structure of the NS (see equations (\ref{eqn:SEa}-\ref{eqn:SEc})). In other words, we ignore the centrifugal bulging that is induced by the initial rotation. This is a reasonable assumption as long the square of the initial angular frequency $\omega_0^2 \ll \omega_k^2$, where $\omega_k$ is the Kepler frequency. Notice that when the star spins up from rest, the background structure is exact within the Newtonian framework that we use. As the star spins up or down, the NS bulges. In addition, interesting elastic stresses arise inside the solid crust. In this section, we develop the mechanical equilibrium equations and boundary conditions for both the fluid and the solid layers within a NS when these elastic stresses are induced. 

In the rest of this work, we assume that the star is spinning around the $\hat{z}$ axis. Moreover, to develop the equations presented in the next sections, we suppose that the initial spin of the star is small ($\omega_0^2 \ll \omega_k^2$). In this way, the centrifugal interaction can be treated as a perturbation to the background star that we describe in section \ref{sec:background}. 

We solve a completely fluid NS to obtain the perturbed gravitational potential. Then, we use that perturbed gravitational potential to satisfy the boundary conditions at the bottom and top of the crust using the finite-element method. Since the crustal mass is roughly 1 \% of the star, we do not expect the crust to influence the perturbed gravitational potential significantly. This is the same approach that references \cite{Franco2000,Fattoyev2018} use. The main advantage of this approach is that it reduces significantly the computation time that we spend on the finite-element simulations. 


\subsection{Notation \label{subsec:notation}}

As we will see, the centrifugal force induces different Lengendre-polynomial modes that do not couple with each other in the fluid layers. The same can be said about centrifugal interactions acting on an axisymmetric solid crust. We use Eulerian perturbations to describe the effects of the change in rotational speed on the NS. We denote Eulerian perturbations with the Greek letter $\delta$. Therefore, the total Eulerian perturbation of a background quantity $G$ is given by
\begin{equation}
    \delta G_{tot}(\vec{r}) = \sum_{l} \delta G_{l}(r) P_l(\text{cos} \ \theta)
    \label{eqn:tot_def}
\end{equation}
where $l$ denotes the Legendre-polynomial mode and $\theta$ is the polar angle.
We use a similar notation to denote any perturbative quantity. For instance, in section \ref{sec:perturbations}, we use $\delta \chi_{tot}$ to represent the centrifugal potential, which can be written as a combination of its $l=0$ and $l=2$ Legendre-polynomial modes. 

Additionally, we represent Lagrangian perturbations by the Greek letter $\Delta$. The Lagrangian perturbation operator $\Delta \equiv \delta + \vec{u} \cdot \vec{\nabla}$, where $\vec{u}(\vec{r})$ represent the small displacement of a material element that is originally located at a position $\vec{r}$ with respect to the center of the NS. The total Lagrangian perturbation of a perturbed quantity $G$ is given by $\Delta G_{tot}$, which can be written in a way analogous to equation \ref{eqn:tot_def}.

We use the Latin sub-indices $i$, $j$, and $k$ to denote the spatial components. We use the Einstein summation convention unless explicitly stated.


\subsection{Completely Fluid Star \label{subsec:fluid}}

We consider a completely fluid star to calculate the perturbed gravitational potential that we use in the finite-element simulations to satisfy the crust-core and crust-ocean interface conditions. We suppose that there is a shallow ocean at the surface of the crust. We assume that the fluid within the star is a perfect fluid. For the completely fluid star, the perturbations against the background star that stem from changes in rotational speed are given by the perturbed Euler and Poisson equations
\begin{equation}
    \nabla_i \delta p_{tot} + \delta \rho_{tot} \nabla_i \Phi + \rho \nabla_i \delta \Phi_{tot} - f_i = 0
\label{eqn:pert_euler}
\end{equation}
and
\begin{equation}
    \nabla^2 \delta \Phi_{tot} = 4 \pi G \delta \rho_{tot},
    \label{eqn:pert_poisson}
\end{equation}
where $f_i = \rho \omega^2_{\text{diff}} s \hat{s}$ is the centrifugal force, $s$ is the distance from the axis of rotation, $\hat{s}$ is the corresponding unit vector, $\omega^2_{\text{diff}} \equiv \omega^2 - \omega_0^2$, and $\omega$ and $\omega_0$ are the final and initial angular frequencies, respectively. The small change in pressure $\delta p_{tot} = c_s^2 \delta \rho_{tot}$.

The centrifugal force can be written as $f_i = - \rho \nabla_i \delta \chi_{tot}$. The centrifugal potential $\delta \chi_{tot} \equiv - \tfrac{1}{2} \omega^2_{\text{diff}} s^2$. We can re-write the potential in spherical coordinates as $\delta \chi_{tot} = \tfrac{1}{2} \omega^2_{\text{diff}} r^2 \text{sin}^2 \theta$, where $r$ is the radial distance from the center of the star and $\theta$ is the polar angle with respect to the z axis (which is the rotational axis). The centrifugal potential can be written in terms of the Legendre polynomials $P_0(\text{cos} \ \theta) = 1 $ and $P_2(\text{cos} \ \theta) = \tfrac{1}{2} (3 \text{cos}^2 \theta - 1)$. The result is $\delta \chi_{tot}(r,\theta) = -\tfrac{1}{3} \omega^2_{\text{diff}} r^2 \times (P_0(\text{cos} \ \theta)) - P_2(\text{cos} \ \theta)) = \delta \chi_0(r) P_0(\text{cos} \ \theta) + \delta \chi_2(r) P_2(\text{cos} \ \theta)$, where $\delta \chi_0(r) = - \delta \chi_2(r) = - \tfrac{1}{3} \omega^2_{\text{diff}} r^2$. We follow Gittins, Andersson, and Jones \cite{Gittins2021} and define $\delta U_{tot} \equiv \delta \Phi_{tot} + \delta \chi_{tot}$. We find that solving the equations with this newly-defined variable is easier. Equations (\ref{eqn:pert_euler}) and (\ref{eqn:pert_poisson}) can be re-written in terms of $\delta U_{tot}$ as 
\begin{equation}
    \nabla_i \delta p_{tot} + \delta \rho_{tot} \nabla_i \Phi + \rho \nabla_i 
    \delta U_{tot} = 0
\label{eqn:pert_euler_2}
\end{equation}
and
\begin{equation}
    \nabla^2 \delta U_{tot} = 4 \pi G \delta \rho_{tot} - 2 \omega^2_{\text{diff}}
    \label{eqn:pert_poisson_2}
\end{equation}
The constant term at the right-hand side arises from the Laplacian of the $l=0$ term of the centrifugal force. We can separate equations (\ref{eqn:pert_euler_2}) and (\ref{eqn:pert_poisson_2}) into decoupled $l=0$ and $l=2$ equations. The radial part of each $l$-th mode of the perturbed Poisson equation (\ref{eqn:pert_poisson_2}) is given by
\begin{equation}
    \delta U_l^{\prime \prime} + \frac{2}{r} \delta U_l^{\prime} - \frac{\beta^2}{r^2} \delta U_l = 4 \pi G \delta \rho_l - 2 \omega^2_{\text{diff}} \delta_{l0} \ ,
    \label{eqn:pert_poisson_l}
\end{equation}
where $\beta^2 \equiv l(l+1)$ and $\delta_{l0}$ is 1 if $l=0$ and 0 otherwise. The angular part of each $l$-th mode of the perturbed Euler equation (\ref{eqn:pert_euler_2}) gives
\begin{equation}
    \delta \rho_l = - \frac{\rho}{c_s^2} \delta U_l \ .
    \label{eqn::pert_density_l}
\end{equation}

We impose regularity at the origin on $\delta \Phi_l$ and $\delta \Phi_l^{\prime}$ ($\delta \Phi_l(r=0)$ = 0 and $\delta \Phi_l^{\prime}(r=0) = 0$). In numerical calculations, the perturbed Poisson equations represented by (\ref{eqn:pert_poisson_l}) are indefinite at the origin. Nonetheless, this can be solved by performing a Taylor expansion around a small value of $r=r_0$:
\begin{equation}
    \delta \Phi_l(r_0) = \sum_k \frac{1}{k!}\delta \Phi_l^{(k)}(0) \ r_0^k \ .
    \label{eqn:BCs_delta_Phi_origin}
\end{equation}
In this equation, $\delta \Phi_l^{(k)} (0)$ represents the $k$-th radial derivative around the origin. In the present study, we consider even perturbations. Consequently, all of the coefficients of odd powers of $r_0$ vanish. Notice that for the $l=0$ mode, the centrifugal interaction scales with $r_0^2$, which implies that the coefficient $\delta \Phi_l^{(0)} (0) = 0$.
At the surface of the star, the perturbed gravitational potential must match the exterior solution. Therefore, the boundary condition at the surface is given by
\begin{equation}
    \delta \Phi_l^{\prime} (R) + \frac{l+1}{R} \delta \Phi_l(R) = 0 \ .
    \label{eqn:BCs_delta_Phi_surface}
\end{equation}
We can re-write the boundary conditions (\ref{eqn:BCs_delta_Phi_origin}) and (\ref{eqn:BCs_delta_Phi_surface}) in terms of $\delta U_l$, $\delta U_l^{\prime}$, and $\chi_l$. The resulting equations are
\begin{equation}
    \delta U_l(r_0) = \sum_k \frac{1}{k!}\delta U_l^{(k)}(0) \ r_0^k \ ,
    \label{eqn:BCs_delta_U_origin}
\end{equation}
\
\begin{equation}
    \delta \chi_0(R) = \frac{1}{3} (R \delta U_0^{\prime}(R) + 3 \delta U_0(R)) \ , 
    \label{eqn:BCs_delta_U_l_0_surface}
\end{equation}
and 
\begin{equation}
    \delta \chi_2(R) = \frac{1}{5} (R \delta U_2^{\prime}(R) + 3 \delta U_2(R)) \ .
    \label{eqn:BCs_delta_U_l_2_surface} 
\end{equation}


\subsection{Solid Crust \label{subsec:crust}}

We consider an anisotropic solid crust under linear elasticity. We assume that the star has an initial spin frequency $\omega_0 \ll \omega_K$ when the solid crust forms under mechanical equilibrium. Furthermore, we suppose that there are no initial elastic stresses, as we pointed out at the beginning of this section. As the NS gradually spins up or down, elastic stresses accumulate within the crust. The perturbed Euler equations for the crust of a spinning-up (-down) NS with induced elastic stresses are given by
\begin{equation}
    \nabla_i \delta p_{tot} - \nabla_j t_{ij} + \delta \rho_{tot} \nabla_i \Phi + \rho \nabla_i \delta \Phi_{tot} - f_i = 0
\label{eqn:pert_euler_crust}
\end{equation}
Typically, a NS changes its rotation rate slowly with time. Hence, we can neglect any time dependence in the equation of the continuity of mass. Therefore, we obtain that the Eulerian perturbation in the density is given by
\begin{equation}
    \delta \rho_{tot} = - \nabla \cdot (\rho \vec{u}) \ ,
    \label{eqn:delta_rho_continuity}
\end{equation}
where $\vec{u}$ is the displacement vector. The elastic stresses are described by the elastic stress tensor
\begin{equation}
    t_{ij} = 2 \mu \bar{\varepsilon}_{ij} + 2 \mu \langle \psi \rangle \bar{\varepsilon}_{xx} \delta_{ix} \delta_{jx} \ ,
    \label{eqn:elastic_stress}
\end{equation}
where $\langle \psi \rangle$ quantifies the degree of anisotropy in the crustal material, and the deviatoric strain $\bar{\varepsilon}_{ij}$ is given by 
\begin{equation}
    \bar{\varepsilon}_{ij} = \varepsilon_{ij} - \frac{1}{3} \delta_{ij} \varepsilon_{kk} \ .
    \label{eqn:strain}
\end{equation}
The strain $\varepsilon_{ij}$ is 
\begin{equation}
    \varepsilon_{ij} = \frac{1}{2} \left( \nabla_i u_j + \nabla_j u_i \right)
\end{equation}
The strain $\varepsilon_{ij}$ includes deformations due to changes in shape and volume, while the deviatoric strain $\bar{\varepsilon}_{ij}$ includes deformations due to changes in shape only, because the strain sourced by the change in volume $1/3 \ I \ \text{tr}(\overleftrightarrow{\varepsilon})$ is subtracted from the total elastic strain $\varepsilon_{ij}$. We assume $\langle \psi \rangle \ll 1$. The anisotropic contribution to the elastic stress given by equation (\ref{eqn:elastic_stress}) makes separation of variables 
very difficult. This is the main reason why we recur to the finite-element method to solve the equations of an anisotropic crust. We can re-write equation (\ref{eqn:pert_euler_crust}) in terms of the stress $\sigma_{ij} \equiv t_{ij} - \delta_{ij} \delta p$ 
\begin{equation}
    - \nabla_j \sigma_{ij} + \delta \rho_{tot} \nabla_i \Phi + \rho \nabla_i \delta \Phi_{tot} - f_i = 0 \ .
    \label{eqn:pert_euler_crust_2}
\end{equation}

The boundary conditions at the bottom and top of the crust are given by the continuity of the traction vector $T_i \equiv \sigma_{ij} \hat{n}_j$, where $\hat{n}$ is the normal to the surface. In our case, the boundary surfaces have a normal $\hat{n} = \hat{r}$. The continuity of the radial component of the traction vector at the fluid-solid interfaces is given by 
\begin{equation}
    \sigma_{ij} \hat{n}_j \hat{r}_i \ |_s = \rho \delta U_{tot} \ |_f \ ,
    \label{eqn:BCs_radial_traction}
\end{equation}
where $\delta U_{tot}$ comes from the completely fluid star that we discuss in sub-section \ref{subsec:fluid}, and the subscripts $f$ and $s$ denote the fluid and the solid layer, respectively. The continuity of the perpendicular traction is also a boundary condition for the interfaces. An inviscid fluid can not support shear stresses. Thus, the boundary condition for the perpendicular component of the traction is given by 
\begin{equation}
    \sigma_{ij} \hat{n}_j \hat{n}_{\perp i} \ |_s = 0 \ |_f \ ,
    \label{eqn:BCs_perp_traction}
\end{equation}
where $\hat{n}_{\perp}$ is a vector that is perpendicular to the normal to the surface one considers. In our case, $\hat{n}_{\perp}$ is perpendicular to $\hat{r}$.

In order to execute the finite-element method for equations (\ref{eqn:pert_euler_crust}-\ref{eqn:BCs_perp_traction}), we need to encapsulate the perturbed Euler equations in the weak form. The variational weak-form formulation is given by
\[
    \int_{\Omega_s} \overleftrightarrow{\sigma}
    (\vec{u}):\overleftrightarrow{\varepsilon}(\delta \vec{u}) \ dV \ + \ \int_{\Omega_s} \delta \rho (\vec{u}) \ \vec{\nabla} \Phi \cdot \delta \vec{u} \ dV \\   
\]
\[
    + \ \int_{\Omega_s} \rho \ \vec{\nabla} \delta \Phi(\vec{u}) \cdot \delta \vec{u} \ dV - \int_{\partial \Omega_s} (\overleftrightarrow{\sigma}(\vec{u}) \cdot \hat{n}) \cdot \delta \vec{u} \ dS \\
\]
\begin{equation}
    = \int_{\Omega_s} \vec{f} \cdot \delta \vec{u} \ dV \ ,
    \label{eqn:pert_Euler_crust_weak}
\end{equation}
where : represents an inner product between two tensors. For example, in Cartesian coordinates, 
\begin{align*}
\sigma : \varepsilon = & \, \sigma_{xx} \varepsilon_{xx} + \sigma_{yy} \varepsilon_{yy} + \sigma_{zz} \varepsilon_{zz} \\
 + \ & \sigma_{xy} \varepsilon_{xy} + \sigma_{xz} \varepsilon_{xz} + \sigma_{yz} \varepsilon_{yz} \\
 + \ & \sigma_{yx} \varepsilon_{yx} + \sigma_{zx} \varepsilon_{zx} + \sigma_{zy} \varepsilon_{zy}
\end{align*}
In the variational formulation, the displacement $\vec{u}$ is the trial function that we want to approximate while $\delta \vec{u}$ is the test function. $\Omega_s$ is the integration domain (the solid crust) and $\partial \Omega_s$ is the sum of all the boundaries (the bottom and top of the crust). We take advantage of reflection symmetries and consider one octave of our spherical NS volume. The number of degrees of freedom (DoF) reduces roughly by a factor of 8. Hence, the memory requirements and computation time reduce by about the same factor. In the spherical octant, $\partial \Omega_s$ includes the points that are at the $x=0$, $y=0$, and $z=0$ boundary planes. Reflection symmetries at these planes imply that $u_x(x=0) = 0$, $u_y(y=0) = 0$, and $u_z(z=0)=0$. These are Dirichlet boundary conditions that we use in our finite-element solver.


\section{Numerical Methods \label{sec:methods}}

We give a concise description of the methods we use to find the solution for the completely fluid star, including both the $l=0$ and the $l=2$ Legendre-polynomial modes. More detailed descriptions of these methods can be found in references \cite{Gittins2021,Kruger2015}. On the other hand, we present a more detailed description of the finite-element method that we use to obtain solutions for the equations of the solid crust, as this is the first time that the finite-element method is used in NS physics.


\subsection{$l=0$ Mode for the Completely Fluid Star \label{subsec:l_0_sol_method}}

For the $l=0$ mode, we use the shooting method to solve equation (\ref{eqn:pert_poisson_l}) subject to the origin and the surface boundary conditions given by equations (\ref{eqn:BCs_delta_U_origin}) and (\ref{eqn:BCs_delta_U_l_0_surface}), respectively. In this particular case, the shooting method consists in guessing the coefficient $\delta U_0^{(2)}(0)$ in equation (\ref{eqn:BCs_delta_U_origin}) with which equation (\ref{eqn:BCs_delta_U_l_0_surface}) is satisfied. To implement the shooting method, we use libraries form Python's 
\texttt{SciPy} package \cite{SciPy2020}, including \texttt{optimize} and \texttt{integrate}. We use the \texttt{solve\_ivp} function from the \texttt{integrate} library to solve the ODE (\ref{eqn:pert_poisson_l}) under the different guesses of $\delta U_0^{(2)}(0)$. We embed this ODE solver within the \texttt{fsolve} function from the \texttt{optimize} library to find the value of $\delta U_0^{(2)}(0)$ that satisfies equation (\ref{eqn:pert_poisson_l}). We use a maximum allowed step-size $h = 10$ cm to solve the ODEs and a relative error of 10$^{-10}$ as our root-finding convergence criterion. Other than that, we use the default parameters of \texttt{solve\_ivp} and \texttt{fsolve}. Note that the default ODE solver of \texttt{solve\_ivp}'s function uses the Dormand-Prince method.


\subsection{$l=2$ Mode for the Completely Fluid Star \label{subsec:l_2_sol_method}}

For the $l=2$ mode, Equation (\ref{eqn:pert_poisson_2}) is a homogeneous ODE, in contrast to the $l=0$ case, which has the in-homogeneous term $-2 \omega^2_{\text{diff}}$. We can take advantage of that homogeneity and use a more computationally-efficient solver.  We define $Z_2(r) \equiv U_2^{\prime}(r)$. We find two linearly-independent solutions using two random guesses of the coefficient $\delta U_2^{(0)}(0)$. We know that the true solution is a linear combination of these two linearly-independent solutions. Therefore, we use the definition of $Z_2(r_0)$ (where $r_0$ is the approximation to the origin, as described in the sub-section \ref{subsec:fluid}) and the boundary condition (\ref{eqn:BCs_delta_U_l_2_surface}) to find the two weights that give the true solution when the two random solution guesses are linearly combined. We use \texttt{SciPy}'s \texttt{solve\_ivp} ODE solver with its default parameters and \texttt{NumPy} \cite{NumPy2020} to perform these endeavors. We set the maximum step-size to be $h = 1.0$ cm when solving the ODE under two random guesses of $\delta U_2^{(0)}(0)$.


\subsection{Finite-Element Method for the Solid Crust \label{subsec:fem_crust}}

\begin{figure}[h!tbp]
    \centering
    \includegraphics[width=0.6\textwidth]{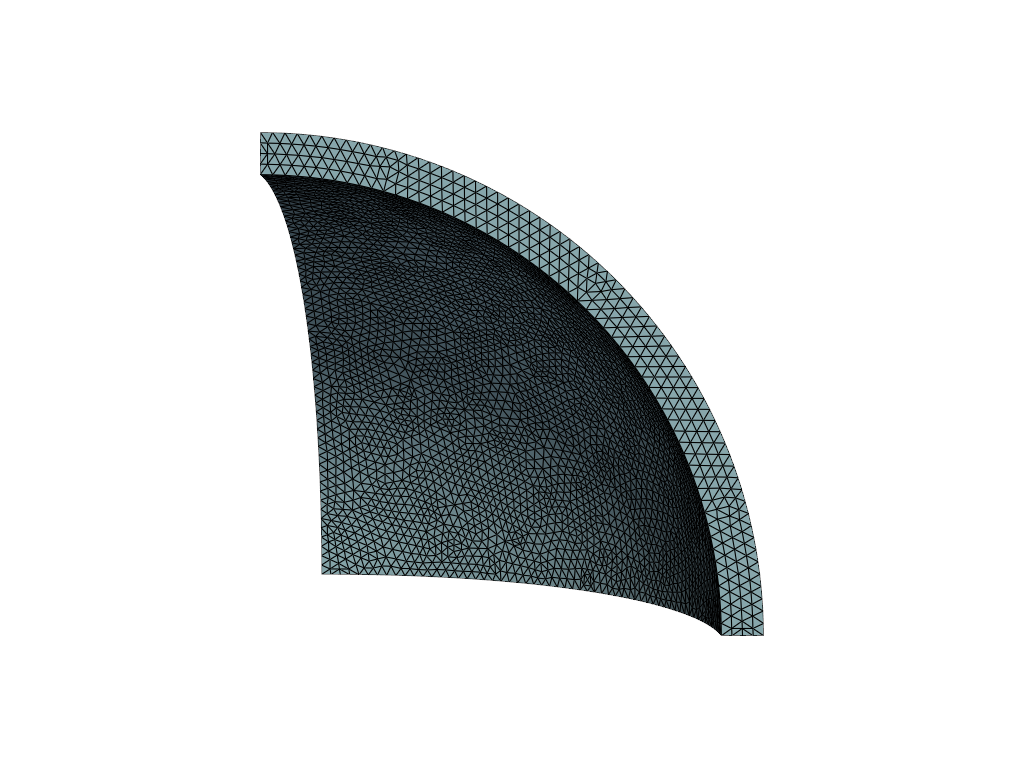}
    \caption{We create a mesh of one octant of the volume of the crust. The elements have an average size of $100$ m, with the exception of elements near the regions where the bottom and top of the solid crust are. In these regions, we make the average mesh size a few times smaller. We produce this three-dimensional plot using \texttt{pyvista}. In this three-dimensional plot, we make the mesh elements much larger to make visualization easier.}
    \label{fig:mesh} 
\end{figure}

As a first step, we use the \texttt{gmsh} software to discretize (mesh) one octant of the background crust volume (see Figure \ref{fig:mesh}). Meshing one octant of the background crust is sufficient to study the displacements, the strains, the stresses, and other physical properties. The rotational deformations of the anisotropic crust under consideration behave the same way in all of the octants because of the reflection symmetries about the $x=0$, $y=0$, and $z=0$. Furthermore, meshing just one octant reduces the number of DoF, the memory consumption, and the computation time by about a factor of 8. In fact, it is this reduction in DoF that allows us to solve the discretized form of equation (\ref{eqn:pert_Euler_crust_weak}) directly, without having to resort to less accurate iterative solution methods. We allow the mesh to have second-order elements that have a maximum size close to $100$ m, and a minimum size that is much smaller than $100$ m. However, the bulk of the elements have sizes close to $100$ m. We force the meshing algorithm to produce elements that are a few times smaller than the maximum allowed size near the bottom and top interfaces ($R_1$ and $R_2$, respectively). The finite-element method solutions are very sensitive to size of the elements, specially to the sizes of the elements near the boundary layers. Hence, we expect that the reduction of element sizes near the solid-fluid interfaces will improve the accuracy of the solutions without incurring in a large computational cost. It is against this mesh that we obtain approximate solutions of the Eulerian displacement $\vec{u}$ and its derived quantities (like the strains, the stresses, the small changes in density, and others). 


After the mesh is crafted, we utilize \texttt{FEniCSx 0.8.0} \cite{Baratta2023,Scroggs2022_2,Scroggs2022,Alnaes2014}, a high-performing and efficient finite-element method library, to solve the discretized form of equation (\ref{eqn:pert_Euler_crust_weak}) for both an anisotropic and an isotropic crust. The solutions for the isotropic and the anisotropic crust, which we call $\vec{u}_0$ and $\vec{u}$, respectively, are found within the mesh elements using second-order Continuous-Garlerkin polynomials. Having $\vec{u}$ and $\vec{u}_0$ on all the nodes of the mesh allows the calculation of the strains, the stresses, the density perturbations, and all of the other physical properties of interest. 

Once we find these solutions, we can calculate the raw ellipticity $e_{\text{raw}}$ that the simulations output. In addition, we can quantify what the ellipticity of an isotropic NS is. We call this the \textit{residual ellipticity}. The residual ellipticity is caused by the imperfect matching of the elements with the boundaries of the sphere octant and other errors associated with the finite-element method. Ideally, the residual ellipticity should be zero because a NS with an isotropic crust that is deformed by rotation is perfectly axisymmetric. 

In order to obtain the raw ellipticity, we calculate the difference in the x-x and y-y components of the moment of inertia tensor across the volume of the NS ($\Omega_{V}$). However, only the volume of the solid crust $\Omega_{s}$ contributes to the ellipticity. Therefore, the difference in the moments of inertia is given by
\begin{equation}
    I_{xx,\text{raw}}-I_{yy,\text{raw}} = \int_{\Omega_{s}} (\rho(r) + \Delta \rho_{tot}(\vec{r})) \ 
    (y^{\prime}(\vec{r})^2 - x^{\prime}(\vec{r})^2) \ \text{d} V \ ,
    \label{eqn:I_x-I_y}
\end{equation}
where $\Delta \rho_{tot} (\vec{r}) = \delta \rho_{tot}(\vec{r}) + \vec{u}(\vec{r}) \cdot \nabla \rho(r)$, $x^{\prime}(\vec{r}) \equiv x + u_x(\vec{r})$ and $y^{\prime}(\vec{r}) \equiv y + u_y(\vec{r})$. In addition, we calculate $I_{zz}$ considering the z-z component of the moment of inertia tensor of the undeformed background star introduced in section \ref{sec:background}:
\begin{equation}
    I_{zz} = \int_{\Omega_{V}} \rho(r) \ 
    (x^2 + y^2) \ \text{d} V \ .
    \label{eqn:I_z}
\end{equation}
This is reasonable because $I_{zz}$ is in the denominator of the ellipticity definition. Consequently, small material displacements and mass density perturbations result in negligible contributions from $\tfrac{1}{I_{zz}}$ to the ellipticity. Notice that the evaluation of equation (\ref{eqn:I_z}) does not require the usage of the finite-element method. We evaluate the expressions in equations (\ref{eqn:I_x-I_y}) and (\ref{eqn:I_z}) in just one octant of the crust, and then calculate the raw ellipticity. This is possible because the evaluation of expressions (\ref{eqn:I_x-I_y}) and (\ref{eqn:I_z}) give the same results for all of the octants. The raw ellipticity is given by 
\begin{equation}
    e_{\text{raw}} \equiv \frac{I_{xx,\text{raw}} - I_{yy,\text{raw}}}{I_{zz}}
    \label{eqn:raw_ellipticity}
\end{equation}

We apply these same steps to calculate the residual ellipticity. The only difference is that we need to evaluate equation (\ref{eqn:I_x-I_y}) with $\vec{u}_0$ replacing $\vec{u}$ because the crust is isotropic ($\langle \psi \rangle = 0$). The difference in moments of inertia for a NS with an isotropic crust is defined as
\[
    I_{xx,\text{raw},\langle \psi \rangle = 0} -I_{yy,\text{raw},\langle \psi \rangle = 0} \equiv I_{xx,\text{raw},0} - I_{yy,\text{raw},0}
\]
\begin{equation}
    = \int_{\Omega_{s}} (\rho(r) + \Delta \rho_{tot,0}(\vec{r})) \ 
    (y^{\prime}_0(\vec{r})^2 - x^{\prime}_0(\vec{r})^2) \ \text{d} V \ ,
    \label{eqn:I_x-I_y_residual}
\end{equation}
where $\Delta \rho_{tot,0} (\vec{r}) = \delta \rho_{tot,0}(\vec{r}) + \vec{u_0}(\vec{r}) \cdot \nabla \rho(r)$, $\delta \rho_{tot,0} = - \nabla \cdot (\rho \vec{u}_0)$, $x_0^{\prime}(\vec{r}) \equiv x + u_{x,0}(\vec{r})$ and $y_0^{\prime}(\vec{r}) \equiv y + u_{y,0}(\vec{r})$. The residual ellipticity is given by 
\begin{equation}
    e_{\text{raw},0} \equiv \frac{I_{xx,\text{raw},0}-I_{yy,\text{raw},0}}{I_{zz}} \ .
    \label{eqn:residual_ellipticity}
\end{equation}
In essence, the residual ellipticity $e_{\text{raw},0}$ is noise from the different components of the finite-element method. Therefore, it is convenient to define the \textit{physical ellipticity} as
\begin{equation}
    e\equiv e_{\text{raw}} - e_{\text{raw},0} \ .
    \label{eqn:physical_ellipticity}
\end{equation}
The physical ellipticity is the ellipticity that stems from crustal anisotropies and changes in rotation.


\section{Results and Discussion \label{sec:results}}

In this section, we present how the results of this work support our previous investigation on anisotropic neutron star crusts and CGW \cite{Morales2023}, and discuss some new findings. 

In our previous work, we modeled the anisotropic crust of a rotating NS as a constant density, rotating thin disc. The simplicity of the problem allowed us to obtain a useful analytical expression for the ellipticity:
\begin{equation}
    e\approx \frac{m_{cr}}{M} \langle \psi \rangle \frac{\Omega^2 - \Omega_0^2}{\Omega_{k,\text{r}}^2} \ .
    \label{eqn:analytical_ellipticity_rel} 
\end{equation}
Here, $m_{cr}$ is the mass of the crust, $M$ is the total mass of the NS, $\langle \psi \rangle$ is the same average anisotropy parameter as in equation (\ref{eqn:elastic_stress}), $\Omega_0$ is the initial rotational frequency, $\Omega$ is the final rotational frequency, and $\Omega_{k,\text{r}} \equiv \omega_{k,\text{r}} / 2 \pi$ is the relativistic Kepler rotational frequency. This ellipticity is the by-product of the anisotropy in the elastic stresses that build within the crust as the star spins up or down. The Kepler rotational frequency for a rotating NS is $\Omega_{k,\text{r}} \approx 1400$ Hz. In this research work, the rotational Kepler frequency is $\Omega_k \equiv \omega_k / 2 \pi \approx 2200$ Hz because we consider a non-relativistic canonical rotating NS with a polytropic EOS. Thus, the analytical estimate that we need to verify becomes
 \begin{equation}    
    e\approx \frac{m_{cr}}{M} \langle \psi \rangle \frac{\Omega^2 - \Omega_0^2}{\Omega_{k}^2} \ .
    \label{eqn:analytical_ellipticity}
 \end{equation}

\begin{figure}[h!tbp]
    \centering
    \includegraphics[width=0.5\textwidth]{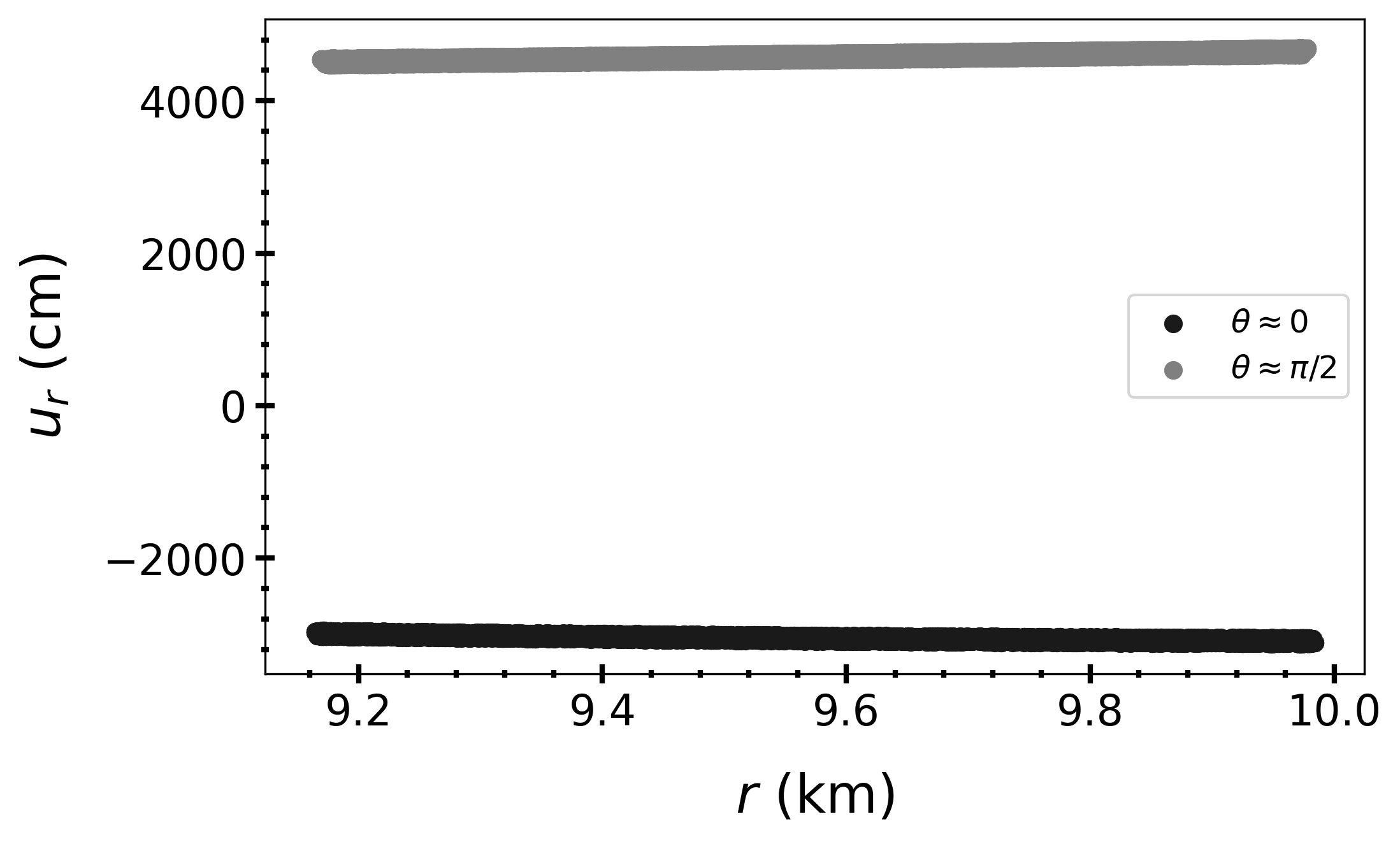}
    \caption{$u_r$ versus $r$ near the pole ($\theta \approx 0$) and near the equator ($\theta \approx \pi/2$). At other positions, $u_r$ has values between the two shown contours.}
    \label{fig:u_r} 
\end{figure}

\begin{figure}[h!tbp]
    \centering
    \includegraphics[width=0.45\textwidth]{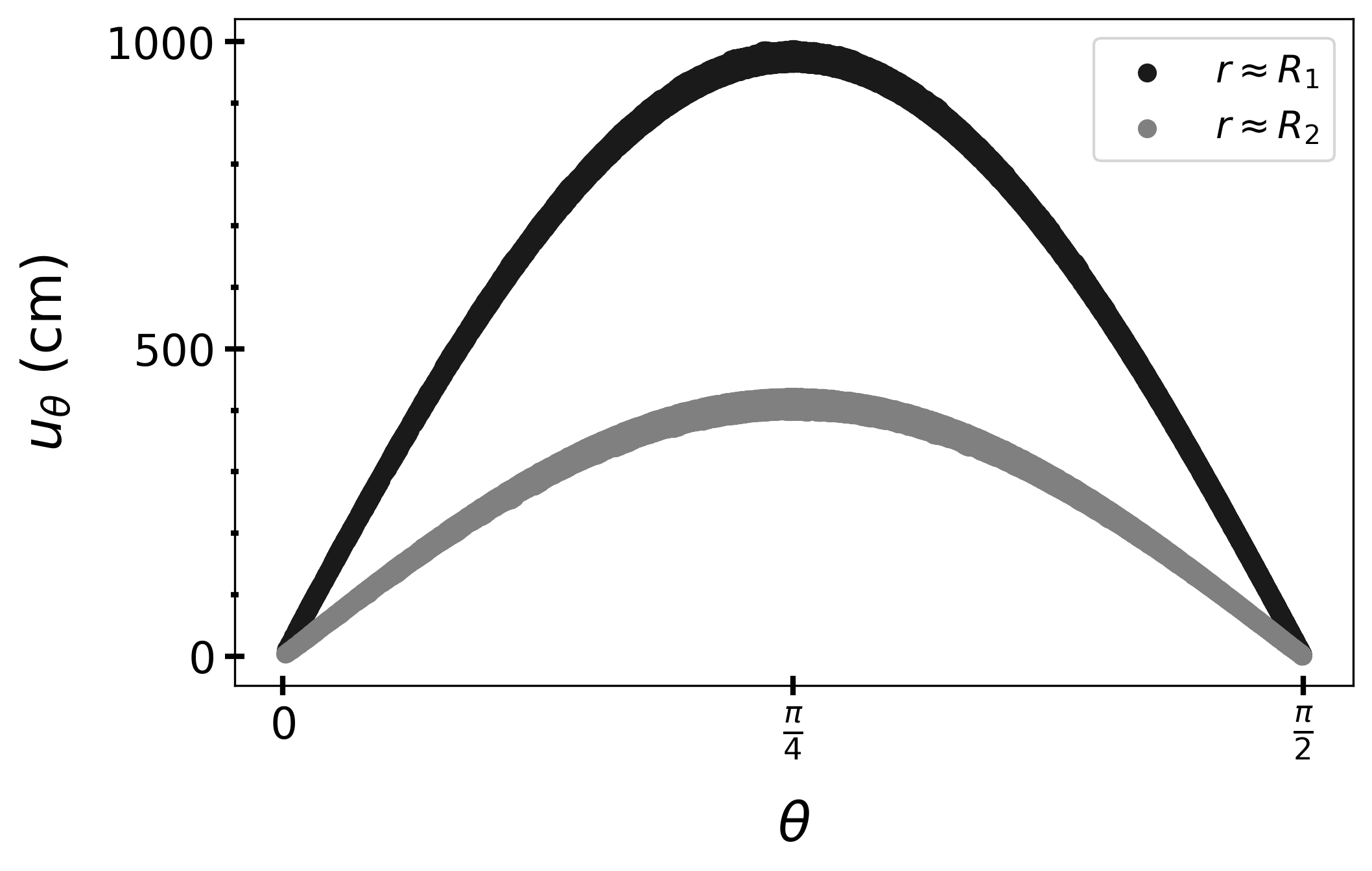}
    \caption{$u_{\theta}$ versus $\theta$ near the crustal bottom ($r \approx R_1$) and near the crustal top ($r \approx R_2$). At other positions, $u_{\theta}$ has values between the two shown contours.}
    \label{fig:u_th} 
\end{figure}

\begin{figure}[h!tbp]
    \centering
    \includegraphics[width=0.45\textwidth]{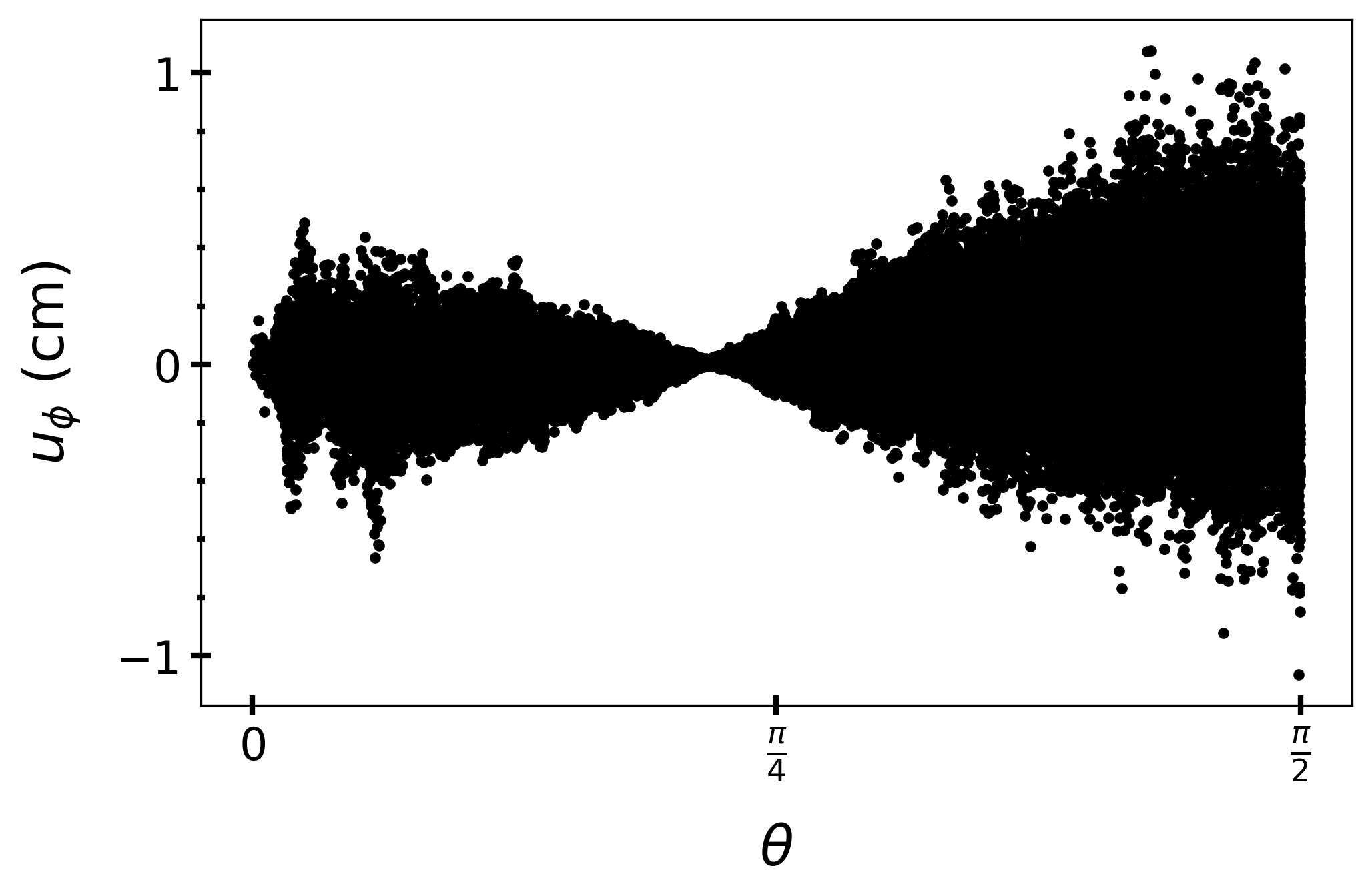}
    \caption{$u_{\phi}$ versus $\theta$.}
    \label{fig:u_ph} 
\end{figure}

For a canonical 1.4 M$_{\odot}$ and 10 km NS, the fraction of the crustal mass is $m_{cr} / M \approx 0.01$. We assume that the NS is initially at rest ($\Omega_0 = 0$ Hz) and spins up to $\sim$ 10 \% of $\Omega_k$ ($\Omega = 220$ Hz). Furthermore, we suppose that the crust has a modest average anisotropy of $\langle \psi \rangle = 10^{-4}$. The finite-element method yields the Eulerian displacement $\vec{u}$ as the solution. The components of the displacement vector in spherical coordinates are shown in Figures \ref{fig:u_r}-\ref{fig:u_ph}. In these figures, $u_r$, $u_{\theta}$, and $u_{\phi}$ represent the displacements in spherical coordinates ($r$,$\theta$,$\phi$), where $r$ is the radial coordinate, $\theta$ is the polar angle, and $\phi$ is the azimuthal angle. We only show the displacement components as functions of the polar angle $\theta$ because that is the strongest dependence. Notice that $u_{\theta} \neq 0$ because of the material anisotropies. 


The evaluation of the analytical estimate of the ellipticity (equation (\ref{eqn:analytical_ellipticity})) gives $e\approx 10^{-2} \times 10^{-4} \times 0.1^2 = 10^{-8}$. The ellipticity that our finite-element application gives (see equations (\ref{eqn:equatorial_ellipticity}), (\ref{eqn:I_x-I_y}), and (\ref{eqn:I_z}))
is $e= 6.2 \times 10^{-9} \sim 10^{-8}$, while the residual ellipticity is $e_0 = 2.2 \times 10^{-10}$. 

\begin{figure}[h!tbp]
    \centering
    \includegraphics[width=0.45\textwidth]{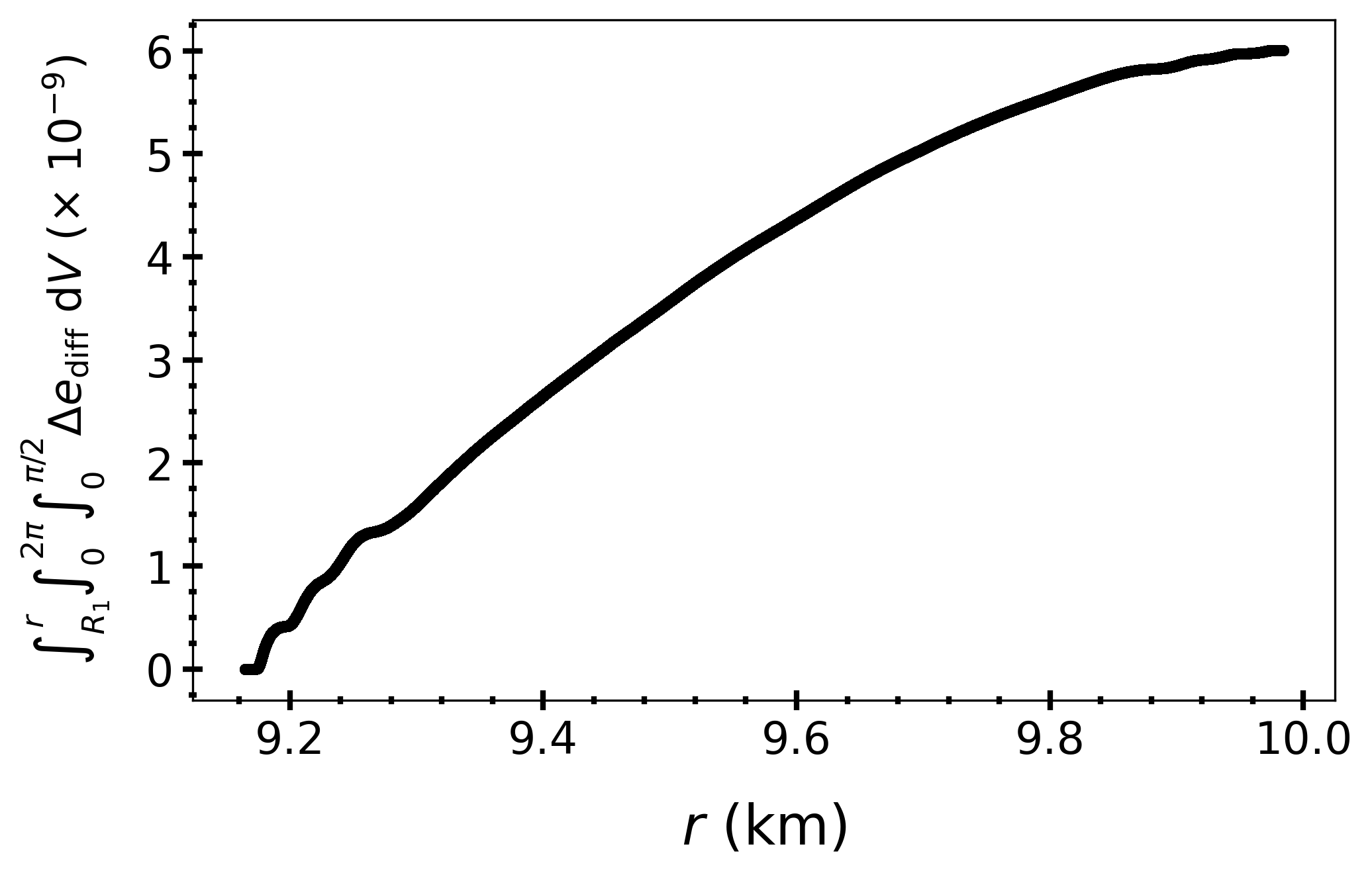}
    \caption{Total ellipticity up to a radius $r$. The value at the crustal top $r = R_2 \approx 10$ km is $\Delta e\equiv e- e_0$, which can be obtained from the values shown in Table \ref{tab:simulation_results}.}
    \label{fig:cumsum_eps} 
\end{figure}

\begin{figure}[h!tbp]
    \centering
    \includegraphics[width=0.45\textwidth]{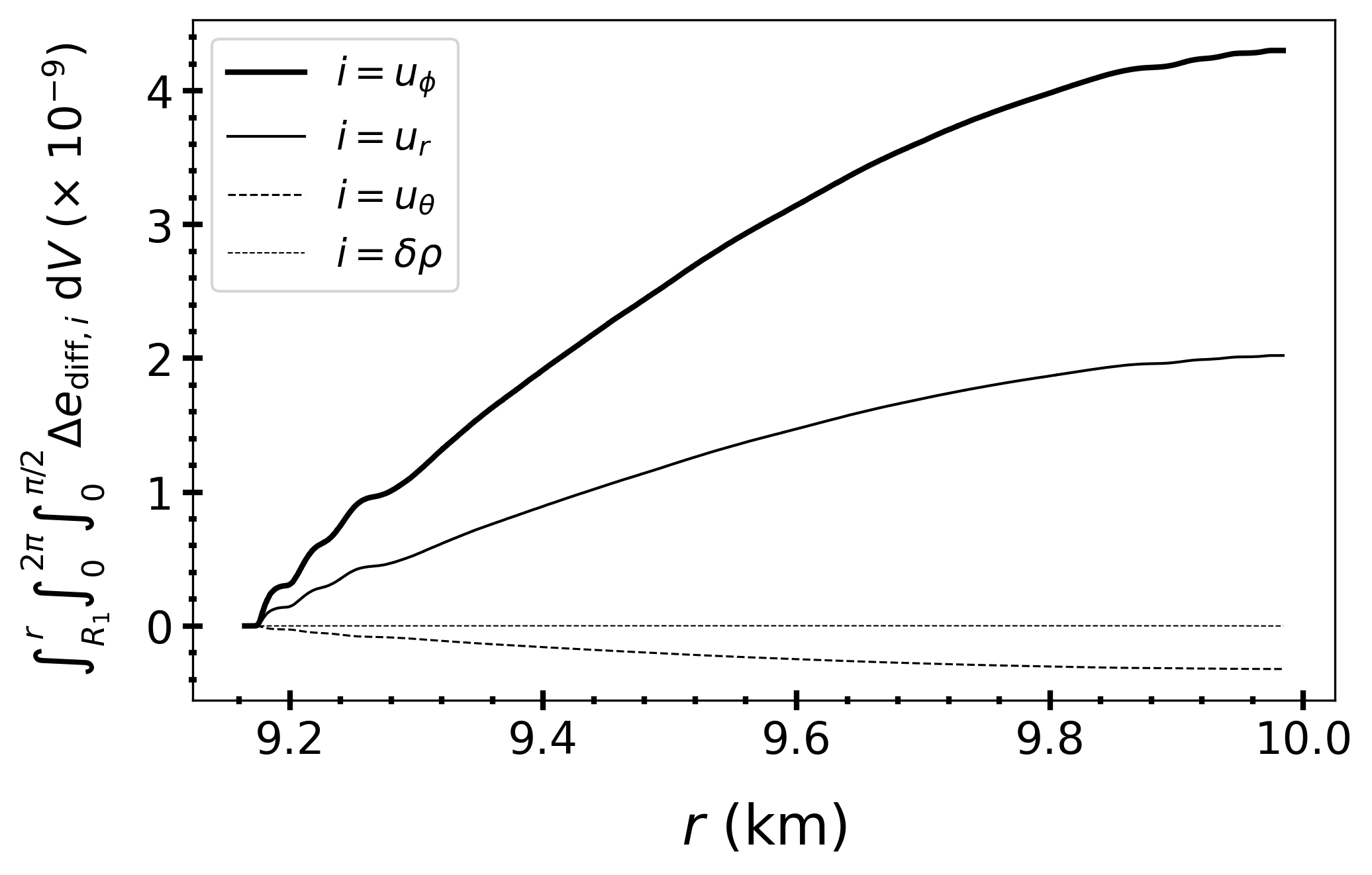}
    \caption{Contributions to the elllipticity up to a radius $r$. The largest contributions to the total ellipticity come from $u_{\phi}$ and $u_r$.}
    \label{fig:cumsum_eps_i} 
\end{figure}

We can study how the ellipticity grows as one goes from the inside of the crust to its outside. Moreover, we can investigate how much the displacement components and the changes in density contribute to that ellipticity growth as the radius gets larger. To resolve these features, it is necessary to subtract the residual ellipticity contribution from the ellipticity contribution on each element. The results are given in Figures \ref{fig:cumsum_eps} and \ref{fig:cumsum_eps_i}.


The integrals have an integration domain that encompasses the whole NS. However, our calculations were done using only one octant of the whole NS. These two calculations are equivalent, as explained above. 

In Figure \ref{fig:cumsum_eps_i} the $\Delta e$'s are the various ellipticity contributions from each element. The diff subscript indicates that we take the difference between the ellipticity contribution from the anisotropic NS (the physical ellipticity) and the perfectly-axisymmetric NS (the residual ellipticity). The formulae from the different $\Delta \varepsilon$'s are given by
\[
    \Delta e_{\text{diff}}(\vec{r}) \equiv (\rho(r) + \Delta \rho_{tot}(\vec{r})) \ 
    (y^{\prime}(\vec{r})^2 - x^{\prime}(\vec{r})^2) \ -
\]
\begin{equation}
    (\Delta \rho_{tot} \longleftrightarrow \Delta \rho_{tot,0}, \vec{r}^{\ \prime} \longleftrightarrow \vec{r}) \ ,
    \label{eqn:cumsum_deps}
\end{equation}
\[
    \Delta e_{\text{diff},u_r}(\vec{r}) \equiv 2 (y^2 - x^2) \left( \frac{u_r(\vec{r})}{r} \rho(r) + u_r(\vec{r}) \rho^{\prime}(r) \right) -
\]
\begin{equation}
    (u_r \longleftrightarrow u_{r,0})
    \label{eqn:cumsum_deps_u_r} \ ,
\end{equation}
\[
    \Delta e_{\text{diff},u_{\theta}}(\vec{r}) \equiv \frac{2 z}{r s} (y^2 - x^2) u_{\theta}(\vec{r}) \rho(r) \ - 
\]
\begin{equation}
    (u_{\theta} \longleftrightarrow u_{\theta,0}) \ , 
    \label{eqn:cumsum_deps_u_th}
\end{equation}
\begin{equation}
    \Delta e_{\text{diff},u_{\phi}}(\vec{r}) \equiv \frac{4 xy}{s} u_{\phi}(\vec{r}) \rho(r) - (u_{\phi} \longleftrightarrow u_{\phi,0}) \ ,
    \label{eqn:cumsum_deps_u_ph}
\end{equation}
\[
    \Delta e_{\text{diff},\delta \rho}(\vec{r}) \equiv \delta \rho_{tot}(\vec{r}) (y^2 - x^2) \ -
\]
\begin{equation}
    (\delta \rho_{tot}\longleftrightarrow \delta \rho_{tot,0}) \ .
    \label{eqn:cumsum_deps_drho}
\end{equation}
Remember that $s$ is the distance from the rotation axis. The total ellipticity's per-element contribution is given by $\Delta e_{\text{diff}} = \Delta e_{\text{diff},u_{\phi}} + \Delta e_{\text{diff},u_r} + \Delta e_{\text{diff},u_{\theta}} + \Delta e_{\text{diff},\delta \rho}$, where terms of second order in $\vec{u}$ are neglected.

Figure \ref{fig:cumsum_eps} seems to disagree with the ellipticity value $e= 6.2 \times 10^{-9}$ that we mentioned above. Nonetheless, this is not a contradiction, as we are subtracting the residual ellipticity contribution from the ellipticity contribution at each element. Since the residual ellipticity if of the order of $10^{-10}$, this difference is expected.

The major contribution to the total ellipticity comes from the $u_{\phi}$ contribution, as we predicted in our previous work \cite{Morales2023}. This is the main reason why our result agrees with our previous work. The second major contribution to the total ellipticity comes from $u_r$. In our previous work, we predicted that the contribution from $u_r$ is much smaller than the contribution from $u_{\phi}$. In contrast, in this work, the $u_{r}$ and $u_{\phi}$ ellipticity contributions are about the same order of magnitude. This difference is due to the density gradient that is taken into account in this work while neglected in our previous work, where we assume that the density is constant through the crust. The $\delta \rho$ and $u_{\theta}$ contributions to the total ellipticity are much smaller. This is another reason why the ellipticity calculation in this work agrees with the estimate that was done in our previous work (see equations  (\ref{eqn:analytical_ellipticity_rel}) and (\ref{eqn:analytical_ellipticity})).

\begin{figure}[h!tbp]
    \centering
    \includegraphics[width=0.45\textwidth]{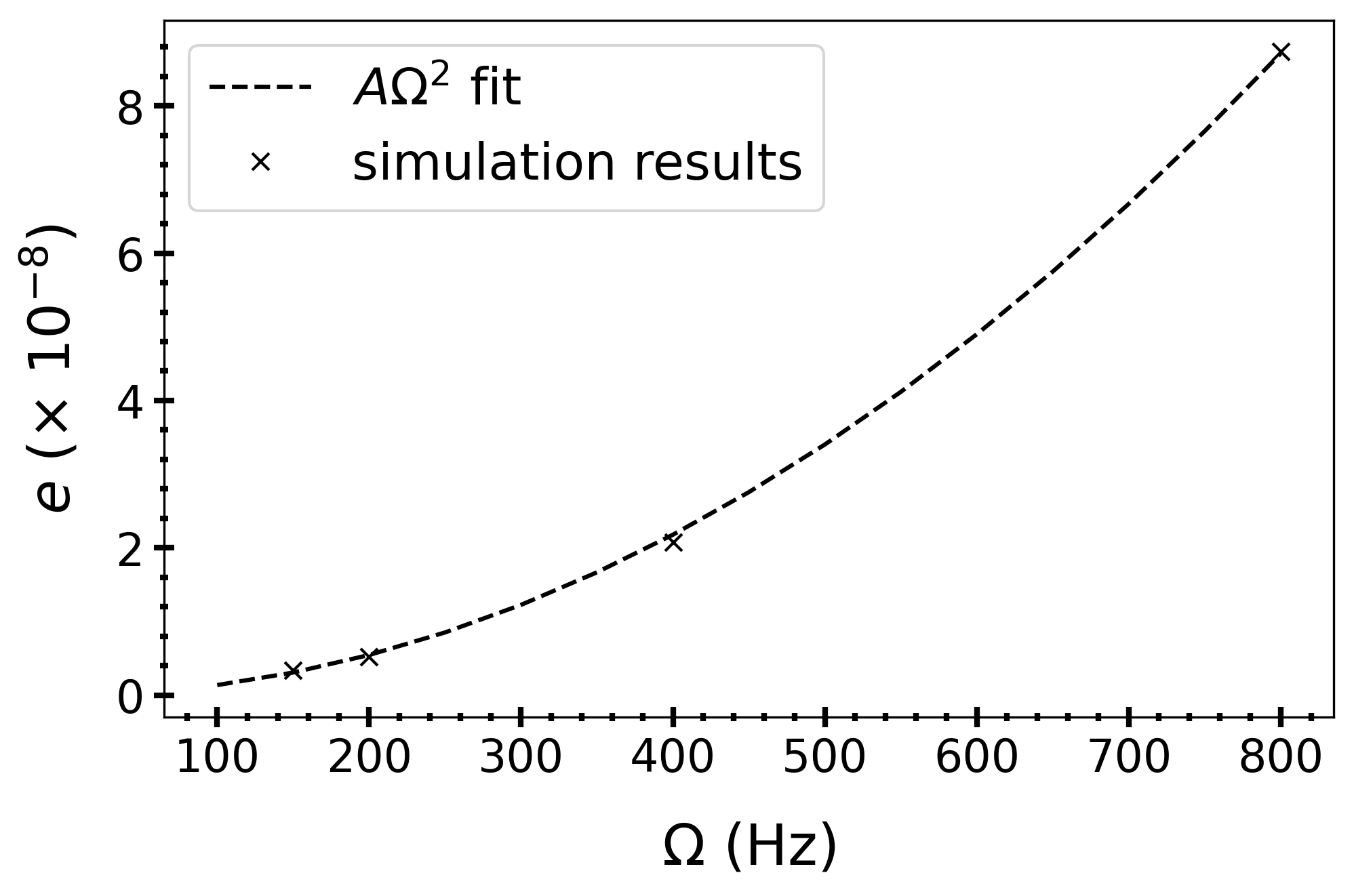}
    \caption{The simulation results, which are denoted by $\times$, agree with a $\Omega^2$ dependence, as shown by the fit and predicted by expressions (\ref{eqn:analytical_ellipticity_rel}) and (\ref{eqn:analytical_ellipticity}). $A=1.36 \times 10^{-13} \ 1/\text{Hz}^2$ is the best-fit parameter.}
    \label{fig:eps_vs_final_Omega} 
\end{figure}

\begin{figure}[h!tbp]
    \centering
    \includegraphics[width=0.45\textwidth]{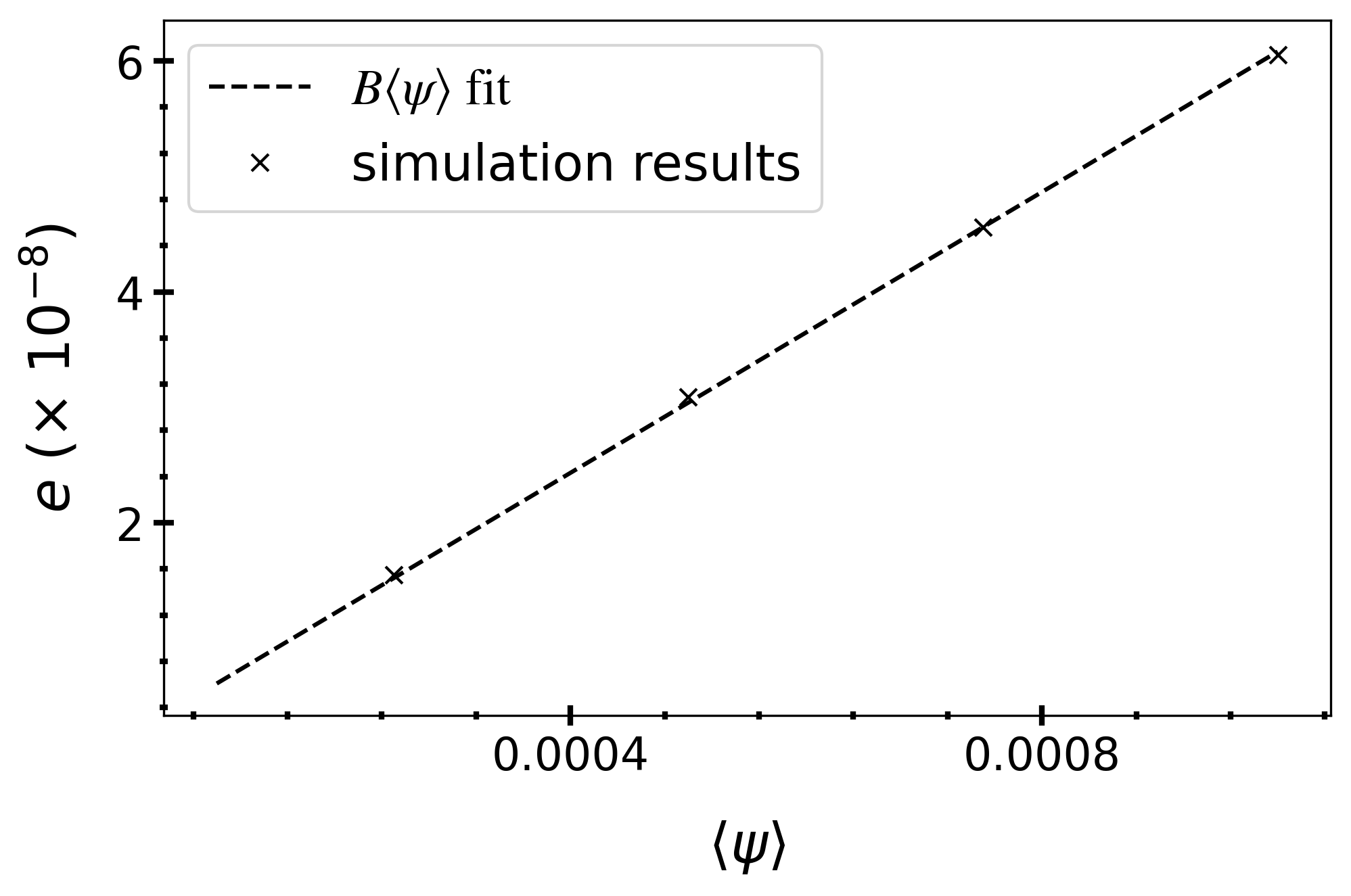}
    \caption{The simulation results, which are denoted by $\times$, agree with a $\langle \psi \rangle$ dependence, as shown by the fit and predicted by expressions (\ref{eqn:analytical_ellipticity_rel}) and (\ref{eqn:analytical_ellipticity}). $B=6.08 \times 10^{-5}$ is the best-fit parameter.}
    \label{fig:eps_vs_phi} 
\end{figure}

In Figures \ref{fig:eps_vs_final_Omega} and \ref{fig:eps_vs_phi}, we confirm that the ellipticity scales linearly with the average anisotropy parameter $\langle \psi \rangle$ and quadratically with $\Omega^2$ for a NS that is initially at rest. This is the expected behavior that we infer from the analytical estimates expressed in equations (\ref{eqn:analytical_ellipticity_rel}) and (\ref{eqn:analytical_ellipticity}). The slight deviation from perfect linearity between the ellipticity and the average anisotropy parameter $\langle \psi \rangle$ comes from the non-linear dependence of the strain on $\langle \psi \rangle$. These non-linearities are negligible when $\langle \psi \rangle \ll 1$. 


Therefore, we can conclude that equations (\ref{eqn:analytical_ellipticity_rel}) and (\ref{eqn:analytical_ellipticity}) are valid analytical estimates of the ellipticity of a slightly anisotropic NS crust that is subject to a small change in rotational frequency. The results of many simulations are given in Table \ref{tab:simulation_results}. This has several consequences that were explained in detail in our previous work \cite{Morales2023}. We briefly summarize these consequences here. 

\addtolength{\tabcolsep}{1pt} 
\begin{table*}[t]
    \begin{tabular}{||c | c c c c c c c||}
        \hline
    Vary & $\Omega$ (Hz) & $\langle \psi \rangle$ & $\rho_{top}$ (g cm$^{-3}$) & element size (m) & $e_{\text{raw},0}$ & $e_{\text{raw}}$ & $e$\\ 
        \hline\hline
    $\Omega$ & 150 & $10^{-4}$ & $10^{12}$ & 100 & $5.5 \times 10^{-10}$ & $3.3 \times 10^{-9}$ & $2.8 \times 10^{-9}$ \\
        & 200 & $10^{-4}$ & $10^{12}$ & 100 & $2.8 \times 10^{-10}$ & $5.2 \times 10^{-9}$ & $4.9 \times 10^{-9}$ \\
        & 400 & $10^{-4}$ & $10^{12}$ & 100 & $9.0 \times 10^{-10}$ & $2.1 \times 10^{-8}$ & $2.0 \times 10^{-8}$ \\
        & 800 & $10^{-4}$ & $10^{12}$ & 100 & $7.8 \times 10^{-9}$ & $8.7 \times 10^{-8}$ & $7.9 \times 10^{-8}$ \\
        \hline
    $\langle \psi \rangle$ & 220 & $10^{-4}$ & $10^{12}$ & 100 & $2.2 \times 10^{-10}$ & $6.2 \times 10^{-9}$ & $6.0 \times 10^{-9}$ \\
        & 220 & $2.5 \times 10^{-4}$ & $10^{12}$ & 100 & $4.8 \times 10^{-10}$ & $1.5 \times 10^{-8}$ & $1.5 \times 10^{-8}$ \\
        & 220 & $5.0 \times 10^{-4}$ & $10^{12}$ & 100 & $8.7 \times 10^{-10}$ & $3.1 \times 10^{-8}$ & $3.0 \times 10^{-8}$ \\
        & 220 & $7.5 \times 10^{-4}$ & $10^{12}$ & 100 & $6.0 \times 10^{-10}$ & $4.6 \times 10^{-8}$ & $4.5 \times 10^{-8}$ \\ 
        & 220 & $10^{-3}$ & $10^{12}$ & 100 & $5.2 \times 10^{-10}$ & $6.1 \times 10^{-8}$ & $6.0 \times 10^{-8}$ \\
        \hline
    $\rho_{top}$ & 220 & $10^{-3}$ & $10^{12}$ & 100 & $5.2 \times 10^{-10}$ & $6.1 \times 10^{-8}$ & $6.0 \times 10^{-8}$ \\
        & 220 & $10^{-3}$ & $5.0 \times 10^{12}$ & 100 & $6.9 \times 10^{-10}$ & $6.1 \times 10^{-8}$ & $6.0 \times 10^{-8}$ \\
        & 220 & $10^{-3}$ & $10^{13}$ & 100 & $5.0 \times 10^{-10}$ & $6.1 \times 10^{-8}$ & $6.1 \times 10^{-8}$ \\
        \hline
    element size & 220 & $10^{-3}$ & $10^{12}$ & 150 & $-3.8 \times 10^{-9}$ & $5.6 \times 10^{-8}$ & $6.0 \times 10^{-8}$ \\
        & 220 & $10^{-3}$ & $10^{12}$ & 125 & $-2.5 \times 10^{-9}$ & $5.7 \times 10^{-8}$ & $6.0 \times 10^{-8}$ \\
        & 220 & $10^{-3}$ & $10^{12}$ & 100 & $5.2 \times 10^{-10}$ & $6.1 \times 10^{-8}$ & $6.0 \times 10^{-8}$ \\
        \hline
        \hline
    \end{tabular}
    \caption{ \label{tab:simulation_results} This table presents the results of the simulations plotted in Figures \ref{fig:eps_vs_final_Omega} and \ref{fig:eps_vs_phi}, as well as other important results. For instance, we show that the ellipticity does not change significantly when we vary the density at which we place the top of the crust. In addition, we show that the ellipticity does not change significantly when we vary the size of the bulk of the elements. Notice that the variation of $\langle \psi \rangle$ induces variations in the residual ellipticity. This is due to the way in which FEniCSx performs computations in parallel, and does not affect the physical ellipticity significantly.}
\end{table*}
\addtolength{\tabcolsep}{1.5pt} 

First, an anistropic crust can produce a detectable ellipticity if a rapidly rotating NS is close-by (see Figure 4 from reference \cite{Abbott2022_2} and compare with the ellipticity of $10^{-9}$ that we obtain from a slighlty anisotropic NS that spins up from rest to 220 Hz). Second, an anisotrpic crust can explain the torque balance that can limit the rotational frequency of accreting milisecond pulsars \cite{Ushomirsky2000}. Third, an anisotropic crust can explain the evidence of a minimum ellipticity observed for a population of milisecond pulsars \cite{Woan2018}. Finally, an anisotropic crust implies that NS mountains have an evolving braking index that can be very different from 5.

Under the approximation of the static background star described in section \ref{sec:background}, we can obtain valid approximations for a spinning down NS, as long as $\Omega_0^2 \ll \Omega_k^2$. If a NS is initially spinning at 220 Hz, spins down to rest, and have an average anisotropy of $\langle \psi \rangle = 10^{-4}$, then the magnitudes of all of the ellipticities and ellipiticity contributions will be the same as the ones presented for the analogous spinning up star that we have presented in this section. The only difference is that the results will have different signs.

\begin{figure}[h!tbp]
    \centering
    \includegraphics[width=0.45\textwidth]{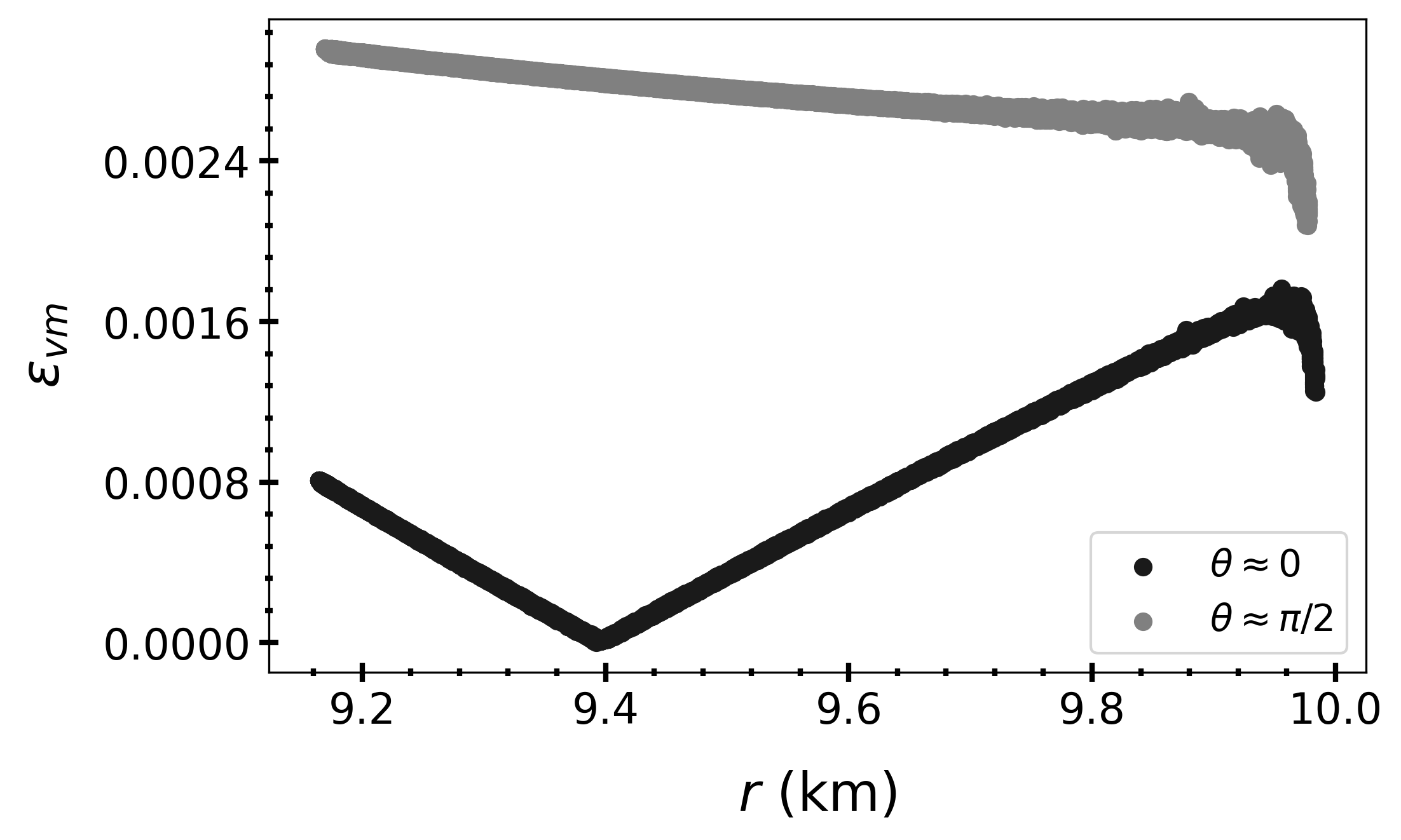}
    \caption{The von-Mises strain versus the radius. The von-Mises strain is largest at the bottom of the crust, at the equator. We show the values of the von-Mises strain at all the radii near the pole and the equator.}
    \label{fig:strain_vm_vs_r} 
\end{figure}

\begin{figure}[h!tbp]
    \centering
    \includegraphics[width=0.45\textwidth]{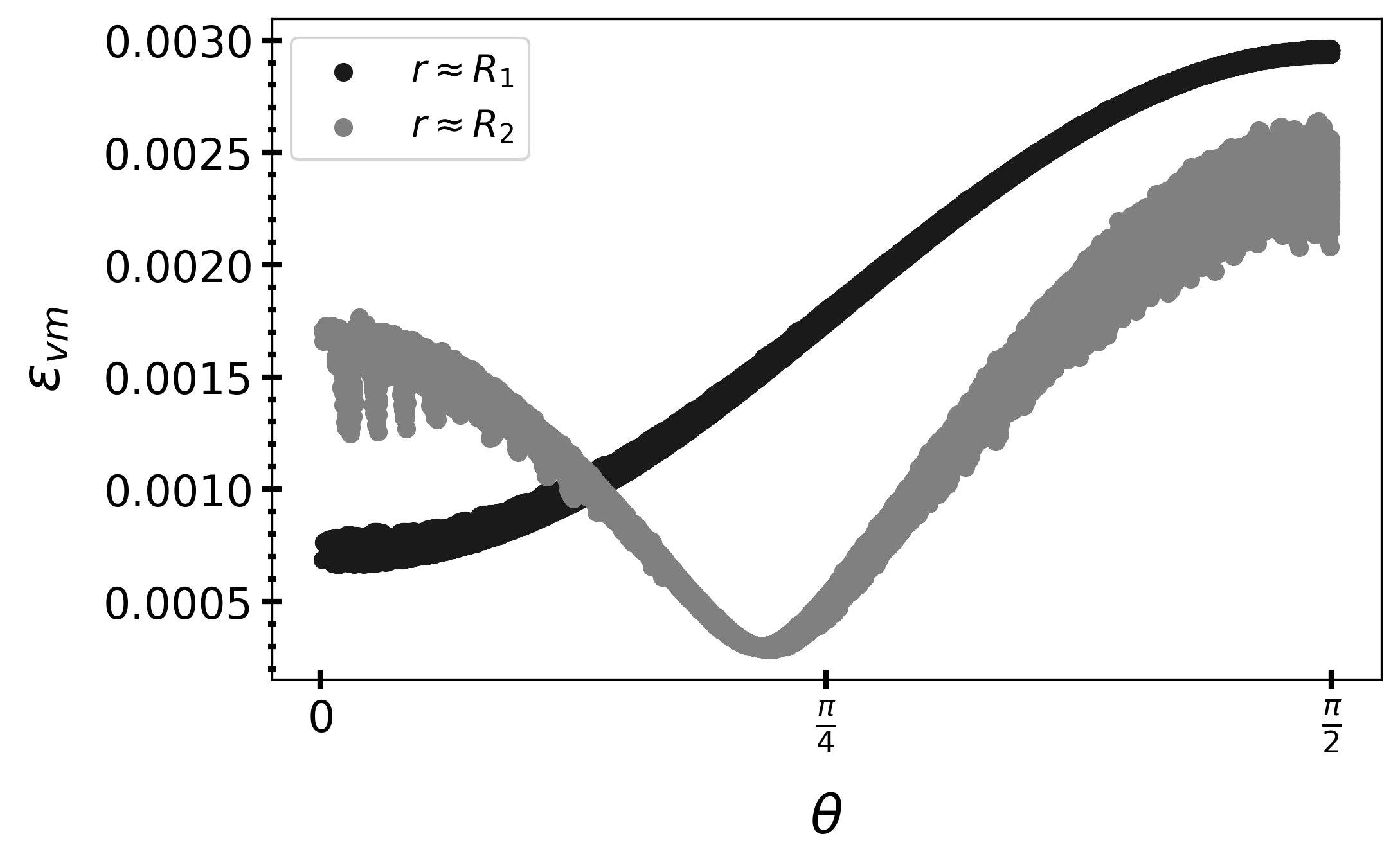}
    \caption{The von-Mises strain versus the polar angle. The von-Mises strain is largest at the equator of the NS. We show the values of the von-Mises strain at all the polar angles near the bottom $R_1$ and top $R_2$ of the crust.}
    \label{fig:strain_vm_vs_th} 
\end{figure}

The von-Mises strain is defined as
\begin{equation}
    \varepsilon_{vm} = \sqrt{ \frac{1}{2} \overleftrightarrow{\bar{\varepsilon}} : \overleftrightarrow{\bar{\varepsilon}}} \ ,
    \label{eqn:strain_vm}
\end{equation}
In spherical coordinates, the von-Mises strain can be written as
\begin{equation}
    \varepsilon_{vm} = \sqrt{ \frac{1}{2} \left( \bar{\varepsilon}_{rr}^2 + \bar{\varepsilon}_{\theta \theta}^2 + \bar{\varepsilon}_{\phi \phi}^2 + 2 (\bar{\varepsilon}_{r r}^2 + \bar{\varepsilon}_{r\phi}^2 + \bar{\varepsilon}_{\theta\phi}^2) \right) }
    \label{eqn:strain_vm_spherical}
\end{equation}
The NS that spins up from rest to 220 Hz has the von-Mises strain field shown in Figures \ref{fig:strain_vm_vs_r} and \ref{fig:strain_vm_vs_th}. The von-Mises strain is largest at the bottom of the crust ($r=R_1$) and at the equator ($\theta=\pi/2$). This is what Franco, Link and Epstein \cite{Franco2000} found. They used a constant-density spheroid with a solid crust atop of a liquid fluid to study quaking NSs. Since $\varepsilon_{vm} \propto \Omega^2 - \Omega_0^2$, the results of our more realistic model support the conclusions of Franco, Link and Epstein \cite{Franco2000} and Fattoyev and Horowitz \cite{Horowitz2009}. Since starquakes start at the crustal bottom, where the crust is densest and strongest, starquakes induced by crust breaking can release sufficient energy to explain some observations, including changes in the pulse profiles (glitches), luminosity increases, and limiting rotational frequencies that are related to crustal strengths. Furthermore, Franco, Link, and Epstein and Horowitz and Fattoyev concluded that crust breaking may cause asymmetric re-distributions of matter that can couple with the rotation of NSs to radiate CGWs. A detailed consideration of crust breaking, starquakes, and observational consequences is left for a future study.


\section{Conclusions \label{sec:conclusions}}

In this research work, we have applied the finite-element method in realistic three-dimensional NS models to study the ellipticities induced by changes in rotational rate coupled with anisotropies in the crust. 

Solid crusts with modest material anisotropies can be stressed by small changes in rotational rate to source ellipticities that are interesting because of various reasons. These ellipticities scale linearly with the small degree of anisotropy in the solid crust and quadratically with the difference between the squares of the final and initial rotational frequecies. Ground-based detectors might be sensitive to these ellipticities if NSs are rapidly spinning and nearby \cite{Abbott2022_2}. Moreover, anisotropic ellipticities can produce the torque balance that may explain accreting millisecond pulsar observations and the minimum ellipticities that might be present in populations of milisecond pulsars \cite{Ushomirsky2000,Fattoyev2018,Morales2023}. 

Our model suggests that the crust breaks at the bottom first. This implies that starquakes that are induced by crust breaking may explain some observations, including glitches, changes in luminosity, limiting rotational frequencies that are related to the strength of the crust, and an asymmetric re-distribution of matter that can couple to the rotation of NSs to radiate CGWs \cite{Franco2000,Fattoyev2018}. These considerations will be studied in detail in future work.



\begin{acknowledgments}
We thank the developers of \texttt{FEniCSx} for their help during the development and implementation of our finite-element simulations. This work is partially supported by the US Department of Energy grant DE-FG02-87ER40365 and National Science Foundation grant PHY-2116686. 
\end{acknowledgments}




\end{document}